# Direct visualization of molecular stacking in quasi-2D hexagonal ice

Yangrui Liu[1,8], Yun Li[2,3,8], Jing Wu[4,8], Xinyu Zhang[2,8], Pengfei Nan[1,*], Pengfei Wang[2,3], Dapeng Sun[5], Yumei Wang[5], Jinlong Zhu[2,3,6,*], Binghui Ge[1,*], Joseph S. Francisco[7,*]

[1] *Information Materials and Intelligent Sensing Laboratory of Anhui Province, Key Laboratory of Structure and Functional Regulation of Hybrid Materials of Ministry of Education, Institutes of Physical Science and Information Technology, Anhui University, Hefei, 230601, China.*

[2] *Shenzhen Key Laboratory of Natural Gas Hydrate, Department of Physics & Academy for Advanced Interdisciplinary Studies, Southern University of Science and Technology, Shenzhen 518055, China*

[3] *Southern Marine Science and Engineering Guangdong Laboratory (Guangzhou), Guangzhou, Guangdong 511458 China*

[4] *Cryo-EM Center, Southern University of Science and Technology, Shenzhen, 518055, China*

[5] *Beijing National Laboratory for Condensed Matter Physics, Institute of Physics, Chinese Academy of Sciences, Beijing, 100190, China.*

[6] *Quantum Science Center of Guangdong-Hong Kong-Macao Greater Bay Area (Guangdong)，Shenzhen 518045, China*

[7] *Department of Earth and Environmental Science, Department of Chemistry, University of Pennsylvania, Philadelphia, PA 19104-6316, USA*

[8] These authors contributed equally: Yangrui Liu, Yun Li, Jing Wu, Xinyu Zhang.

✉ **e-mail:** npf@ahu.edu.cn; zhujl@sustech.edu.cn; bhge@ahu.edu.cn; frjoseph@sas.upenn.edu

**Abstract**

The structure and properties of water or ice are of great interest to researchers due to their importance in the biological, cryopreservation and environmental fields. Hexagonal ice ($I_h$) is a common ice phase in nature and has been extensively studied; however, microstructural investigations at the atomic or molecular scale are still lacking. In this paper, the fine structure of quasi-2-dimensional ice $I_h$ films was directly examined using cryogenic transmission electron microscopy. Two types of thin $I_h$ films were observed: perfect single crystals growing along the [0001] direction and crystals with stacking faults, including both basal (BSF) and prismatic (PSF) ones, along the orientation of [11$\bar{2}$0]; these results were further confirmed by theoretical calculations. Importantly, for the first time, the stacking faults in ice $I_h$ were directly visualized and resolved. In light of the extension behavior of the chair conformation composed of the water molecules, we elucidated the formation mechanism of BSF in $I_h$, namely, the $I_c$ phase. This study not only determined the structural characteristics of ice structure at the molecular scale but also provided important concepts for researchers to more fully understand the growth kinetics of ice crystals at the atomic scale.

**Introduction**

$H_2O$, as a simple but important substance, is treated as the starting point of life on Earth or even in the universe. There are more than 20 crystalline polymorphs of ice [1] due to its stacking diversity, especially the ordering/disordering states of H atoms, which are governed by the Pauling model [2]. These states generally exist and need to be acknowledged. More importantly, the thermal dynamic and kinetic pathway from $H_2O$ freezing to ice is fundamentally critical to understand H₂O-related natural phenomena, geomechanics, biocryopreservation techniques and even the exploration of outer space, such as in comet clouds and interstellar grains with harsh pressure and temperature conditions. Furthermore, these processes need to be microscopically investigated as an equilibrium state or dynamic processes.

It has been reported that liquid $H_2O$ initially crystallizes into H-disordered hexagonal ice ($I_h$), with the metastable H-ordered cubic ice ($I_c$) [3-5] simultaneously formed when the temperature decreases at ambient pressure. Most of the metastable cubic phase can quickly transfer into the $I_h$, while a residual fraction of the cubic phase may survive as basal stacking faults (SFs) between $I_h$ grains. The fraction of $I_c$ is determined by the kinetic energy barrier and its kinetic pathway. For instance, by proper kinetic pathway control, Huang et al. identified a series of $I_c$ defects at the molecular level, revealing the long-standing debate regarding whether $H_2O$ could freeze to form $I_c$ [6]. The first molecular-scale images of nano ice particles were captured by Kobayashi [7] through high-resolution transmission electron microscopy. The stacking faults on $I_h$ and $I_c$ simulated by Carignano [8] were laterally elaborated by Malkin [9] and Kalita [10] through X-ray diffraction experiments, but direct visualization was still challenging.

Stacking faults in similar structures, such as clays, metals, zeolites, and especially those in the wurtzite structure of GaN, have generally been observed [11-14]. With the rapid development of cryoelectronic microscope techniques [6,15-21], the stacking faults in H₂O polymorphs can potentially be quenched and observed. A potential approach to capturing the formation and evolution of the stacking faults at the grain boundary between $I_h$ and $I_c$ is to quench a frame of the dynamic process and visualize it at ultralow temperatures to obtain insights into the stacking fault configuration, such as its density, distribution, orientation, and migration. On the other hand, theoretical calculations can further simulate the energy of the quenched meta-stable state to quantitively describe the crystal growth and transformation between $I_h$ and $I_c$ and to even capture the kinetic process. This is critically important for both cosmic $H_2O$ exploration and bio-cryopreservation techniques.

In this work, a thin film $I_h$ obtained by cooling liquid $H_2O$ with liquid nitrogen was investigated through cryo-transmission electron microscopy (TEM) with low-dose imaging methods [15,16]. The thin film grown along the hexagonal [0001] axis was a perfect crystal, free of defects, while the one grown along the [11$\bar{2}$0] axis exhibited numerous stacking faults, which were confirmed by theoretical simulations. Furthermore, the formation of BSF, in the form of $I_c$, was elucidated by means of the chair conformation of $H_2O$ molecules.

**Results**

Quasi 2-dimensional (2D) thin film of ice was distributed on the TEM grid, as shown in Fig. 1a. To determine the films' phases, electron diffraction was carried out, and two distinct patterns were obtained (Figs. 1b and c). These patterns were indexed as the [0001] and [11$\bar{2}$0] zone axes of ice $I_h$ (space group P6$_3$cm, cell parameters: a = b = 7.82 Å, c = 7.36 Å, α = β = 90°, γ = 120° [22]), respectively. Contrary to the flawless diffraction pattern (Fig. 1b), which indicated excellent crystalline quality of the [0001] film, stripes between the diffraction spots were observed from the bottom to the top in the diffraction pattern of [11$\bar{2}$0] (Fig. 1c), showing the presence of planar defects in the film [23].

To resolve the microstructure, especially the defects, high-resolution TEM (HRTEM) images were recorded. In Fig. 1d, a uniform contrast throughout the image was observed, indicating a perfect crystal of the film along the [0001] direction. In contrast, Fig. 1g shows a linear contrast pattern; this indicated the presence of planar defects distributed across the entire film with the [11$\bar{2}$0] orientation. The corresponding fast Fourier transformations (FFTs) inset bottom right in Figs. 1 d and g confirmed these deductions. Corresponding high-magnification images taken from well-crystallized areas are shown in Figs. 1e and h. HRTEM image simulation based on the multislice method [24] was performed using Mactempas [25], with ice $I_h$ as a model. The simulated images (inserted in Figs. 2e and h) effectively matched the experimental results (Figs. 2e and h), confirming that the ice films in this study were mainly of the ice $I_h$ phase, as opposed to the $I_c$ phase reported by Jianguo et al. [21].

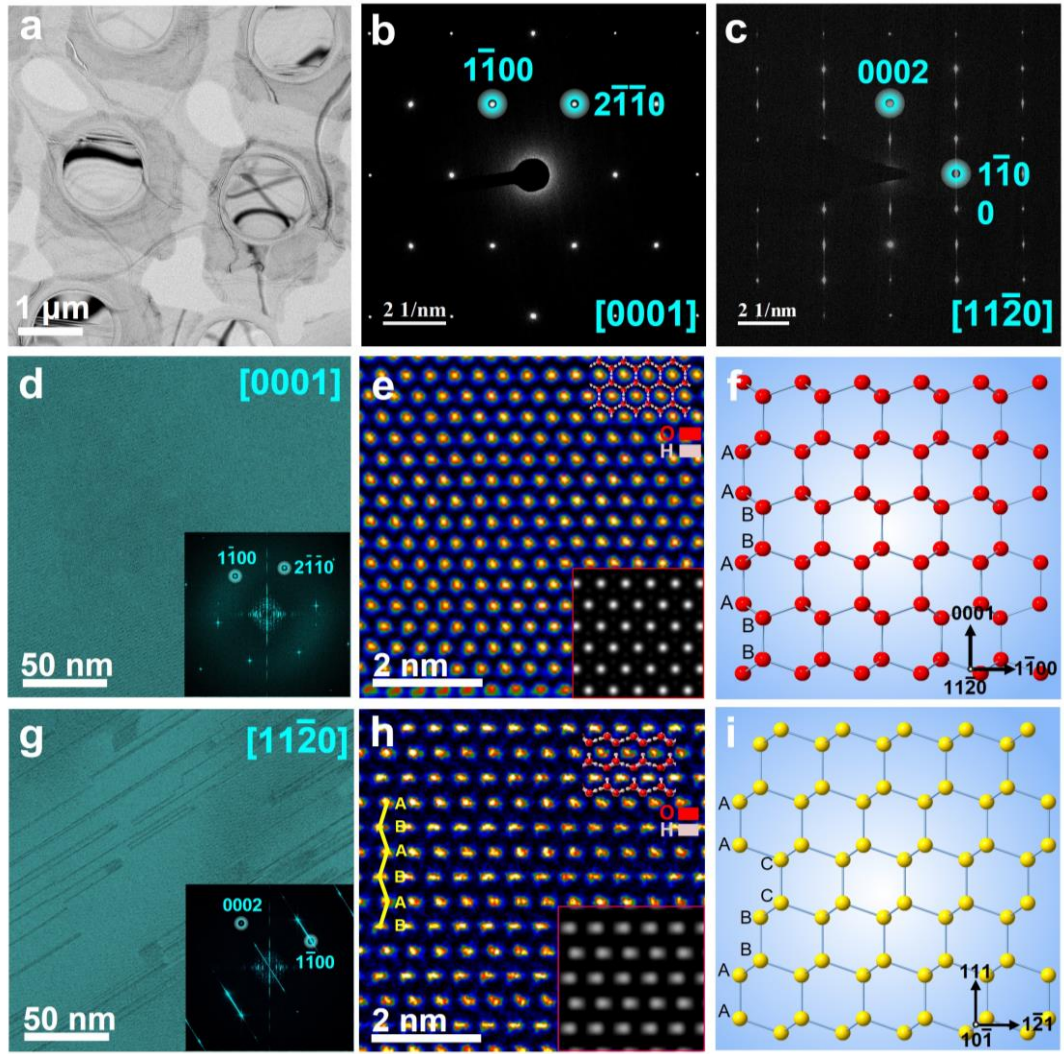

**Fig. 1. Visualization of hexagonal ice (I$_h$). a**, Low magnification TEM image of the ice thin film. **b-c**, Electron diffraction patterns along the [0001] and [11$\bar{2}$0] directions, respectively. **d** and **g**, [0001] and [11$\bar{2}$0] TEM images of ice with the corresponding FFT inset. **e**, [0001] filtered HRTEM image, and the projected atomic model is superimposed in top right corner. And the simulated HRTEM image of ice I$_h$ in the [0001] direction (thickness of 54.5 nm, defocus of -71 nm) is inserted in bottom right corner. **h**, [11$\bar{2}$0] filtered HRTEM image of ice, and the atomic model is superimposed in top right corner. Simulated HRTEM image of ice I$_h$ in the [11$\bar{2}$0] direction (thickness of 76.4 nm, defocus of -31 nm) is inserted in bottom right corner. **f** and **i**, Projected structural models of ice I$_h$ along the orientation of [11$\bar{2}$0] and I$_c$ along the orientation of [10$\bar{1}$], with only oxygen atoms displayed in red and yellow, respectively. A, B and C show the corresponding molecular layers with certain displacements.

Ice $I_h$ belongs to the hexagonal wurtzite lattice, which can be considered a stacked layer along the *c*-axis direction with the (0001) plane as the structural unit, as shown in Fig. 2f. In hexagonal and cubic crystalline systems, the letters A, B and C are commonly used to denote the different layers [26]. Thus, the order of layers in ice $I_h$ (Fig. 2f) can be described as ⋯AABBAABBAABB⋯, where A and B represent two adjacent layers that are staggered by the $\frac{1}{3}[\bar{1}100]$ vector in the basal plane, and position C remains unoccupied. In contrast, in ice $I_c$, as shown in Fig. 2i, the order of layers is ⋯AABBCCAABBCC⋯. From the inset model in Fig. 2h, the bright spots can be observed between the bilayers. To simplify the discussion, A is used to denote bilayer A; specifically, A denotes where the bright spot in Fig. 2h is located, while B and C are similarly defined. In this case, the order of layers in $I_h$ and $I_c$ can be abbreviated as ⋯ABAB⋯ and ⋯ABCABC⋯, respectively.

According to the description of Hirth and Lothe et al. [26], the misalignment of atomic layers is prone to occur along the [0001] orientation in hexagonal structures, which can lead to the formation of stacking faults. These faults can result in stripes in the diffraction patterns as well as the FFT. Frank et al. proposed possible types of stacking faults, which essentially originated from the slip, insertion or withdrawal of atomic layers in the basal plane or the prismatic plane, leading to the formation of BSF and PSF [27].

The HRTEM images were taken in the defective area of the thin film, as shown in Fig. 2. In Fig. 2a, the yellow curve is introduced to highlight the arrangement of bright spots in the experimental image; a break in the stacking order was observed in the middle position. If the stacking sequence was named ···ABAB··· below the defect, then above that, the stacking sequence shifted to ···BCBC···. The change in the stacking sequence indicated $\frac{1}{3}[\bar{1}100]$ slip between the upper and lower sides (Fig. 3b), which corresponded to the BSF in hexagonal crystal systems [27]. As shown in Fig. 2j, the ABC segment could be regarded as the unit of the $I_c$ structure. By comparing the experimental image with the simulated image inset on the right of Fig. 3a, the observed defect could be identified as an $I_1$-type BSF.

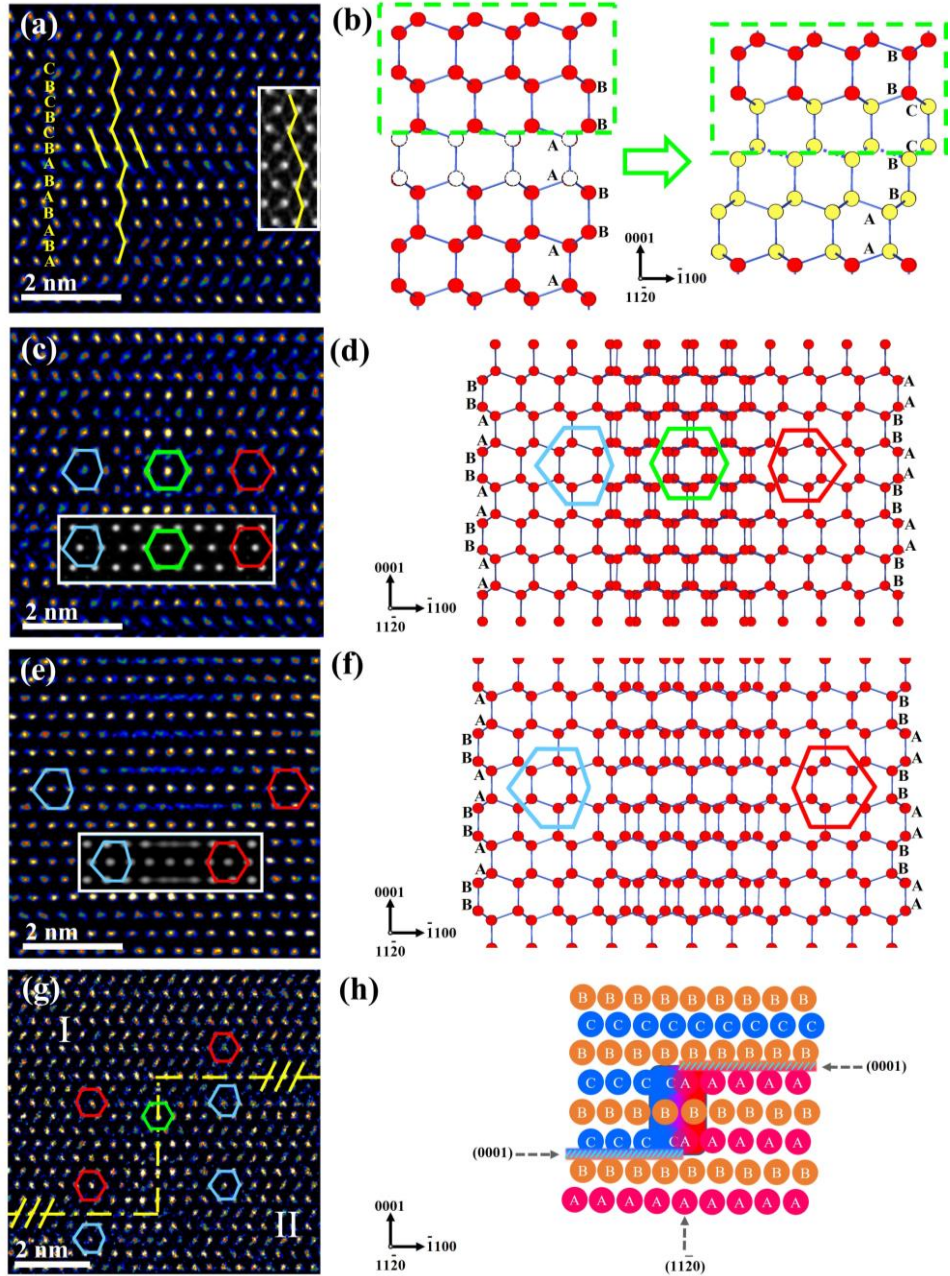

**Fig. 2. Stacking faults in ice I$_h$. a**, **c**, **e** and **g,** Filtered image of the stacking faults observed along the [11$\bar{2}$0] direction in ice I$_h$ with corresponding simulation inset. Yellow curves in (a) show the distribution of bright spots. Blue and red distorted hexagonal rings in (c), (e) and (g) show the feature of the bright spots in the well-crystallized area, and the green hexagon shows the feature of the defective region of PSF. **b**, **d** and **f**, Projected structural models of BSF, PSF 1 and PSF 2, where the yellow circle in (b) represents the cubic unit formed by the defect area. **h**, Schematic diagram of the interlacing BSF and PSF.

In addition to BSFs in the wurtzite structure, as shown in Fig. 2b, Drum [28] and

Blank [29] et al. reported the presence of PSFs characterized by displacive vectors of $\frac{1}{2}[10\bar{1}1]$ and $\frac{1}{6}[20\bar{2}3]$, which were also observed in our experiment. As shown in Figs. 2c and e, the lattice arrangements displayed mirror symmetry, as indicated by the distorted blue and red hexagonal rings on two sides, respectively. Moreover, the arrangement of bright spots in the middle of Fig. 2c closely resembled a perfect hexagon, as highlighted by the green hexagonal ring. In contrast, stripe contrast was observed in Fig. 2e along the horizontal direction. The corresponding simulated images inset in Figs. 2c and 2e with fault vectors $\frac{1}{2}[10\bar{1}1]$ and $\frac{1}{6}[\bar{2}023]$ reported by Drum [28] (named PSF 1) and Blank [29] (named PSF 2) were in agreement with the experimental results, thereby validating the defect model. To the best of our knowledge, stacking faults in ice $I_h$ were directly visualized at the molecular scale for the first time.

The $[11\bar{2}0]$ films were full of stacking faults, as demonstrated in Fig. 1g, and the convergence of two BSFs with different basal planes led to a PSF, as shown in Fig. 2g. A yellow dotted line divided the image into two parts, I and II. The opposite direction of the red and blue hexagons in Parts I and II could be observed, showing that the arrangement of the water molecules in ice $I_h$ changed from ···ABAB··· to ···CBCB··· after crossing the yellow line, as depicted in Fig. 2h.

To quantitively understand the growth of ice $I_h$ and $I_c$, we performed density functional theory (DFT) calculations to determine the formation energies of $I_h$, $I_c$ and the two types of stacking faults (BSF-$I_1$, PSF 1 and PSF 2), as shown in Fig. 3a. In agreement with previous reports, ice $I_h$ was the most stable phase, while ice $I_c$ (XIc-$I4_1mc$) was the second most metastable phase after $I_h$, possessing a marginally higher thermodynamic energy [30-32]. Contrary to the continuous and tight arrangement of hexahedral units in the ice $I_h$ and ice $I_c$ configurations (Fig. 3b), the presence of layer faults was clearly observed in the BSF and PSF phases. Notably, the two PSF configurations exhibited the highest thermodynamic energy, potentially due to the existence of a nonhexahedral polymorph along the [0001] direction (Fig. S1).

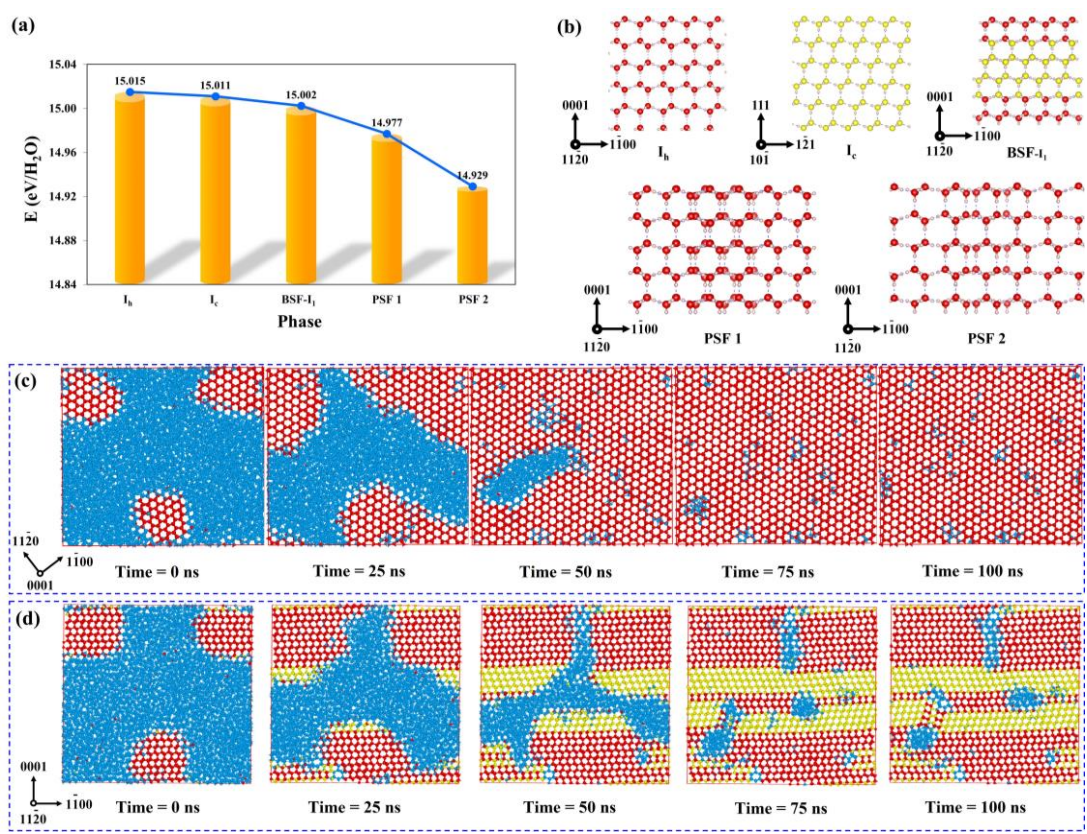

**Fig. 3. Density functional theory calculations for different models. a** and **b**, Formation energies of $I_h$, $I_c$, BSF-$I_1$, PSF 1 and PSF 2 from the DFT calculations and their atomic configurations. Snapshots of the configuration of the $I_h$ growth along the [0001] (c) and [11$\bar{2}$0] (d) zone axes at different simulation time scales from the MD calculations. The sizes of the squares are 9.08 nm × 10.25 nm and 10.50 nm × 10.00 nm along [0001] and [11$\bar{2}$0] orientation, respectively. The red balls, yellow balls, and blue balls represent the oxygen atoms of water molecules in the $I_h$, $I_c$ and liquid phases, respectively.

The kinetic processes involved in the formation and migration of the BSF and PSF

were investigated through simulations. Detailed information regarding the simulation models and methods can be found in the molecular dynamic (MD) calculation methods. The CHILL+ algorithm was extended to categorize each water molecule as $I_h$, $I_c$, interfacial ice, or liquid based on the hydrogen bond arrangement [33]. The progress of ice layer formation at 249 K across various time scales are shown in Figs. 3c and 3d. Visualization of the formation and migration processes is provided in supplemental videos 1 and 2. The evolution of potential energy for the two simulation systems indicated that the potential energy decreased as the ice grew (Fig. S2). The simulations showed the formation of a uniform $I_h$ thin film along the [0001] direction. Simultaneously, a part of the $I_c$ phase, BSF, formed in the $I_h$ substrate along the [11$\bar{2}$0] direction, as shown by the snapshots in Fig. 3d. Interestingly, the $I_c$ phase initially appeared to be predominantly located at the box boundary in Fig. 3d and videos 2; this result indicated that the spatial confinement could modulate the growth of $I_c$, which was reasonable considering the similar forming energies of $I_h$ and $I_c$ (Fig. 3a). Moreover, the other fluctuations could also stabilize $I_c$ (BSF) in the center free part of the simulation box; these fluctuations included pressure fluctuation in the simulation and the temperature gradient, inhomogeneity, and impurity in the real case. Snapshots of ice growth along the [0001] and [11$\bar{2}$0] directions could also be reflected from the evolution of the ice numbers $I_h$ and $I_c$ in these two simulation systems (Fig. S3). Specifically, only the $I_h$ phase was formed along the [0001] direction and ultimately grew into perfect single crystals, while a mixture of $I_h$ and $I_c$, in the form of a stacking fault, was formed along the [11$\bar{2}$0] direction.

The atomic configuration of the BSF and PSF could be explained by examining the arrangement of water molecules in the ice film. The nucleation and growth processes of the 2D ice thin film along the [0001] and [11$\bar{2}$0] directions were both based on the extension of the chair conformation (composed of 6 oxygen atoms in blue in Fig. 4a), as indicated by the literature [34,35]. In this conformation, the oxygen atoms of $H_2O$ molecules were alternately arranged on the bottom and top positions. During the growth process of the (0001) plane, i.e., the extension of the chair conformation along the [1$\bar{1}$00] and [11$\bar{2}$0] directions (Fig. 4a), the two-dimensional translation of the chair conformation was limited due to the sharing of two oxygen atoms between adjacent chair conformations. Therefore, the chair conformation extended in the (0001) plane to form a defect-free lattice, as shown in Fig. 4a. This finding was consistent with the experimental observations (Fig. 1d) and MD simulations (Fig. 4b). However, when the chairs grew along the [0001] direction, two extension pathways for the chair conformation exist because there are no shared oxygen atoms; these pathways are the translational extension and mirror extension (Figs. 4c and d). According to the DFT calculations, the formation energies of hexagonal ice (mirror extension) and cubic ice (translation extension) did not significantly differ, and thus, both extensions could occur. Consequently, during the extension of the chair conformation along the [0001] direction (Fig. 4d), the BSF could form due to the inclusion of both translational and mirror extension, as observed in our experiments (Fig. 1g and Fig. 2) and the MD simulations along the [11$\bar{2}$0] orientation (Fig. 4e). Correspondingly, PSFs were created due to the connection of the (0001) planes formed by different crystal nuclei during the growth process.

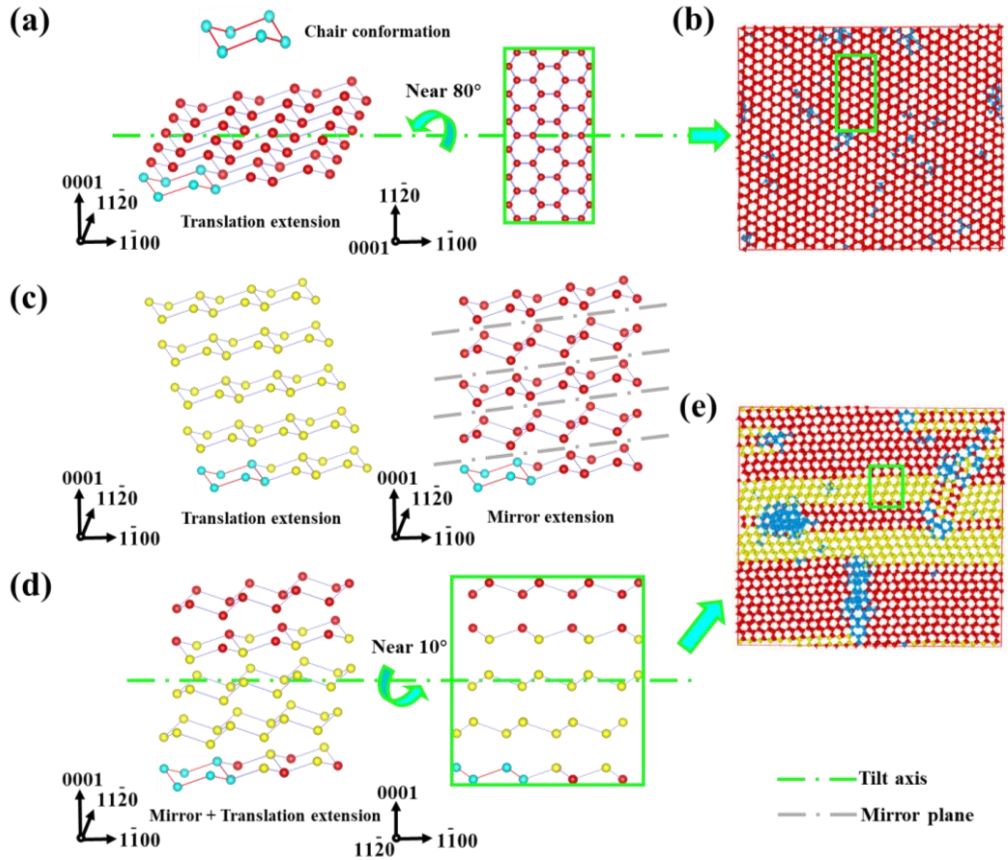

**Fig. 4. The nucleation and growth processes of the 2D ice thin film. a**, Schematic diagram of the translational extension of the chair conformation within the (0001) plane. **b**, Snapshots of $I_h$ along the [0001] projection obtained from the MD simulation. **c**, Schematic diagram of translational extension ($I_c$) and mirror extension ($I_h$) of the chair conformation along the [0001] orientation. **d**, Schematic diagram of translational + mirror extension of the chair conformation within the ($11\bar{2}0$) plane. **e**, Snapshots of $I_h$ along the [$11\bar{2}0$] projection obtained from the MD simulation.

**Discussion**

In summary, quasi 2D thin films of ice crystallized at liquid nitrogen temperature and ice $I_h$ were identified via cryo-HRTEM imaging. Two distinct thin films were obtained: one with the orientation [0001] that was perfectly crystalline, and the other with the orientation [$11\bar{2}0$] that contained numerous stacking faults. By combining the image simulation and DFT calculations, BSF and PSF common to wurtzite structures were confirmed. Our study marks the first time that the stacking faults in ice $I_h$ were directly visualized. Moreover, the possible growth path of the $I_c$ phase, in the form of BSF, was thoroughly demonstrated by the MD simulations; these simulations showed that the phase was stabilized with the translation extension of the chair conformation along the [0001] direction.

**Methods**

**Experiments:** The samples for cryo-TEM observation were carried out by Titan Krios G3i (D3845) microscope equipped with a Schottky X-FEG field emission electron gun and a spherical aberration corrector at 300 kV. All HRTEM experiments were operated at liquid nitrogen temperature and at a pressure of $10^{-5}$ Pa.

**Sample preparation:** Similar to traditional sample preparation for cryo-TEM, samples in this work were prepared using the Vitrobot device from Thermo Fisher Scientific Inc., with the only difference being that liquid nitrogen was used as the freezing agent instead of liquid ethane. The detailed sample preparation process as follows: a volume of 3-5 μL of water is pipetted onto the TEM sample microgrid, and the excess water is blotted off with filter paper after the water is spread out on the microgrid. And then the microgrid is quickly immersed in a copper cup filled with liquid nitrogen for rapid cooling and crystallization. Once crystallization is complete, the microgrid is placed on a sample stage cooled with liquid nitrogen and inserted into the electron microscope column for characterization.

**DFT calculation:** The underlying energy calculations and structure optimizations were performed in the framework of DFT using the Vienna ab initio Simulation Package [36]. The projector-augmented plane-wave [37] approach was used to represent the ion-electron interaction with $1s^2$ and $2s^2 2p^4$ valence electrons for H and O, respectively.

The electron exchange-correlation functional was treated using the generalized gradient approximation proposed by Perdew, Burke and Ernzerhof [38]. The van der Waals (vdW) interactions are also considered using the DFT-D2 vdW corrections. A cutoff energy of 800 eV for the expansion of the wave function into plane-waves and appropriate regular Monkhorst-Pack k-point grids of $2\pi \times 0.05 \text{Å}^{-1}$ were chosen to ensure that all enthalpy calculations were well converged to 1 meV/atom.

**MD calculation:** The initial configuration of the ice/water two-phase system was consistent with 8872 water molecules. The box size of the initial simulation model of Fig. S3a and Fig. S3b were $9.08 \times 10.25 \times 2.91$ nm$^3$ and $2.70 \times 10.50 \times 10.00$ nm$^3$, respectively. GROMACS 2018 software package was employed in the molecular dynamics simulations [39]. The widely used TIP4P/Ice water model was applied for ice nucleation [40]. The SETTLE algorithm was used to constrain the rigidity of water molecules. The melting point of TIP4P/Ice water model was 270 ± 3 K, which was close to the experimental value. Lorentz-Berthelot mixing rules were applied to calculate the Lennard-Jones potentials between different atoms. The leapfrog algorithm was introduced to the equations of motion with a time step of 1 fs [41]. The long-range electrostatic interactions were evaluated by the Particle Mesh Ewald method with a cutoff of 1 nm [42]. The short-range interactions were truncated at 1 nm. The energy minimization with the steepest descent algorithm was firstly performed in all simulations. Then, a 200 ps (constant volume) *NVT* ensemble was performed to relax the initial configuration at 249 K. Subsequently, the molecular dynamics simulation was performed under (isothermal-isobaric) *NpT* ensemble at 1 bar and 249 K for 100 ns for the simulation model that expose the [0001] planes of ice I$_h$. For the simulation model that exposes the [11$\bar{2}$0] planes of ice I$_h$, the molecular dynamics simulation was performed for 1000 ns with the same ensemble. Nosé-Hoover thermostat [43] (with a relaxation time of 1 ps) and Parrinello-Rahman barostat [44] (with a relaxation time of 4 ps) were applied to control the temperature and pressure, respectively. Periodic boundary condition was performed in *x, y,* and *z* directions [45].

**Data availability**

The data generated and analysed in this study are included with the paper.

## Acknowledgments

This work is supported by the National Key R&D Program of China (Grant No. 2022YFA1403203), the National Natural Science Foundation of China (Grant Nos. 11874394, 11775011), the University Synergy Innovation Program of Anhui Province (No. GXXT-2020-003), Shenzhen Key Laboratory of Natural Gas Hydrates (No. ZDSYS20200421111201738), Southern Marine Science and Engineering Guangdong Laboratory (Guangzhou, Grant No. K19313901), Laboratory of Natural Clathrates

Platform, Academy for Advanced Interdisciplinary Studies (SUSTech). The work was also supported by Center for Computational Science and Engineering at Southern University of Science and Technology, and supported by the Major Science and Technology Infrastructure Project of Material Genome Big-science Facilities Platform supported by Municipal Development and Reform Com-mission of Shenzhen.

**Author contributions**

P.N., B.G. and J.Z. supervised the project. Y.L., P.N., J.W., P.S. and Y.W. performed the experiment. Y.L., J.W., P.W. and X.Z. performed the computational calculations. J.Z., B.G and J.F. advised on the theoretical modelling. Y.L., P.N. and Y.L. wrote the manuscript.

**Competing interests**

The authors declare no competing interests.

**Additional information**

**Correspondence and requests for materials** should be addressed to Pengfei Nan, Jinlong Zhu, Binghui Ge or Joseph S. Francisco.

**Online content**

Any methods, additional references, Nature Portfolio reporting summaries, source data, extended data, supplementary information, acknowledgements, peer review information; details of author contributions and competing interests; and statements of data and code availability are available at https:xxxxxxxxx